%% file: root.tex
\documentclass[letterpaper, 10 pt, conference]{ieeeconf}  % Comment this line out if you need a4paper

\IEEEoverridecommandlockouts                              % This command is only needed if 
\usepackage{graphics} % for pdf, bitmapped graphics files
\usepackage{epsfig} % for postscript graphics files
\usepackage{mathptmx} % assumes new font selection scheme installed
\usepackage{times} % assumes new font selection scheme installed
\usepackage{amsmath,amssymb,amsfonts}
\usepackage{subfigure}
\usepackage{array}
\usepackage{algorithm}
\usepackage{algpseudocode}
\usepackage{subfigmat}
\usepackage{rotating}
\usepackage{enumerate}
\usepackage{adjustbox}
\usepackage{paralist}
\usepackage{hyperref}
\hypersetup{
    colorlinks=true,
    linkcolor=black,
    filecolor=black,      
    urlcolor=black,
    citecolor=black,
    bookmarks=true,
    pdftitle={The impact of modeling approaches on controlling safety-critical, highly perturbed systems: the case for data-driven models},
    }

\newcommand{\mb}[1]{{\ensuremath{\mathbf{#1}}}}   % math bold face
\newcommand{\Rbb}[1]{\mathbb{R}^{#1}}		    % R^n
\newcommand{\vect}[1]{\mb{#1}} % vector symbol: non-italic - this template does not have bold math

\newtheorem{rep@theorem}{\rep@title}
\newcommand{\newreptheorem}[2]{%
\newenvironment{rep#1}[1]{%
 \def\rep@title{#2 \ref{##1}}%
 \begin{rep@theorem}}%
 {\end{rep@theorem}}}
\makeatother

\newreptheorem{theorem}{Theorem}

\newreptheorem{lemma}{Lemma}

\usepackage{nicefrac}
\newcommand{\onehalf}{\nicefrac{1}{2}}

\usepackage{subfigure}
\usepackage{subfigmat}

\graphicspath{{img/}}

\title{\LARGE \bf
The impact of modeling approaches on controlling safety-critical, highly perturbed systems: the case for data-driven models.}

\author{Piotr Łaszkiewicz$^{1, 2}$, Maria Carvalho$^{3}$,  Cláudia Soares$^{2}$ and Pedro Lourenço$^{4}$% <-this % stops a space
\thanks{$^{1}$GNC Division, Flight Segment and Robotics unit of GMV, Poland {\tt\small plaszkiewicz@gmv.com}}%
\thanks{$^{2}$Nova School of Science and Technology, U. Nova de Lisboa {\tt\small claudia.soares@fct.unl.pt} {\tt\small p.laszkiewicz@campus.fct.unl.pt}}%
\thanks{$^{3}$Boeing Commercial Airplanes, Frankfurt am Main, Germany {\tt\small maria.c.carvalho@boeing.com}}%
\thanks{$^{4}$GNC Division, Flight Segment and Robotics unit of GMV, Portugal {\tt\small palourenco@gmv.com}}%
}

\begin{document}

\maketitle
\thispagestyle{empty}
\pagestyle{empty}

%%%%%%%%%%%%%%%%%%%%%%%%%%%%%%%%%%%%%%%%%%%%%%%%%%%%%%%%%%%%%%%%%%%%%%%%%%%%%%%%
\begin{abstract}
    This paper evaluates the impact of three system models on the reference trajectory tracking error of the LQR optimal controller,
    in the challenging problem of guidance and control of the state of a system under strong perturbations and reconfiguration.
    We compared a smooth Linear Time Variant system learned from data (DD-LTV) with state of the art Linear Time Variant (LTV) system identification methods,
    showing its superiority in the task of state propagation. Moreover, we have found that DD-LTV allows for better performance
    in terms of trajectory tracking error than the standard solutions of a Linear Time Invariant (LTI) system model,
    and comparable performance to a linearized Linear Time Variant (L-LTV) system model. We tested the three approaches on the perturbed and time varying spring-mass-damper systems.

    % To solve the problem of identification of discrete-time linear time-varying systems from data, we propose a closed-form method that derives the exact and optimal solution, formulating the learning problem as a regularized least squares problem where the regularizer favors smooth solutions within a trajectory. The algorithm has a complexity that increases linearly with the number of samples considered per sampled trajectory and it outputs the desired results even in the presence of large volumes of data. To show its applicability to real world systems, we test with a spring-mass-damper system and use the estimated model to find the optimal control path, and then test the same procedure in simulators for a Space mission that requires precise pointing of the on-board cameras in a fast dynamics environment. This paper provides a fast alternative to classical system identification techniques for linear time-varying systems.

\end{abstract}

\section{Introduction}
% Structure
% paragraph 1: Broad problem
% paragraph 2: more specific
% paragraph 3: more more specific
% paragraph 4: more more more specific
% paragraph 5: Wrap up the story
% When citing, refer to the relevant contribution to our story 
%
%
% paragraph 1: Broad problem
% First sentence: every reader of the paper should understand this sentence. Make it compelling. State the broad problem. Move from broad to the narrow.

\begin{figure}[!htb]
    \begin{subfigmatrix}{3}
        \subfigure[Continuous reconfiguration of a system. The system parameters $p$ change continuously in time.\label{fig:cont-reconfiguration}]
        {\includegraphics[width=0.45\linewidth]{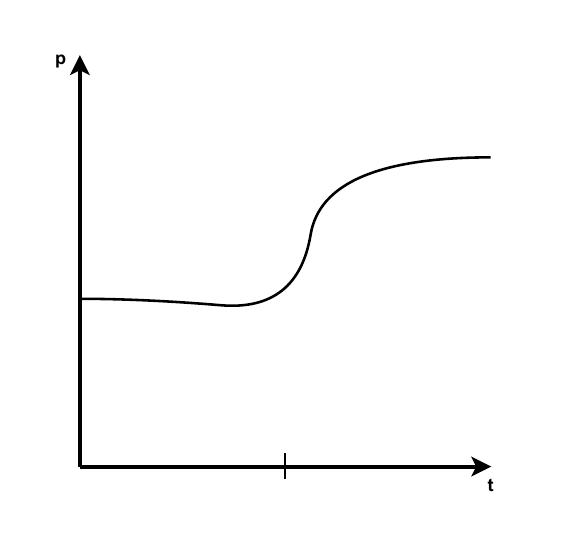}}
        \subfigure[Instantaneous reconfiguration of a system. The system parameters $p$ change discontinuously in time.\label{fig:inst-reconfiguration}]
        {\includegraphics[width=0.45\linewidth]{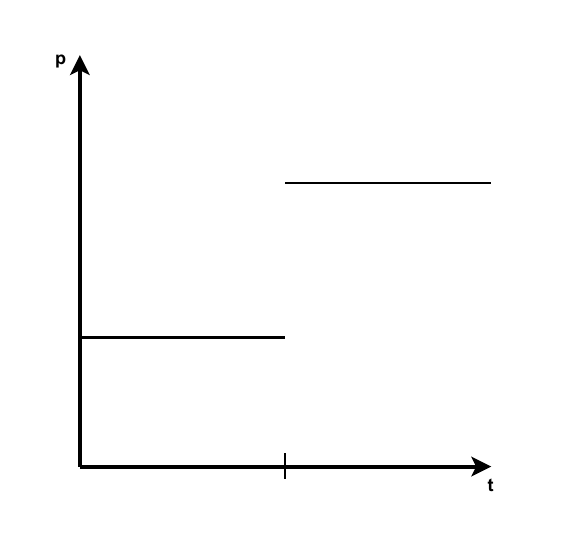}}
        \subfigure[Variation of linearized error dynamics between consecutive samples for a spring-mass-damper system undergoing mixed reconfiguration.
        $\vect{C}(k)$ represents concatenation of system matrices $\vect{A}(k)$ and $\vect{B}(k)$, defined in section \ref{sec:COSMIC}.
        It is clear that the variations are bounded from one instant to the next, entailing smoother transitions.\label{fig:variation}]
        {\includegraphics[width=\linewidth, trim=25 35 45 55,clip]{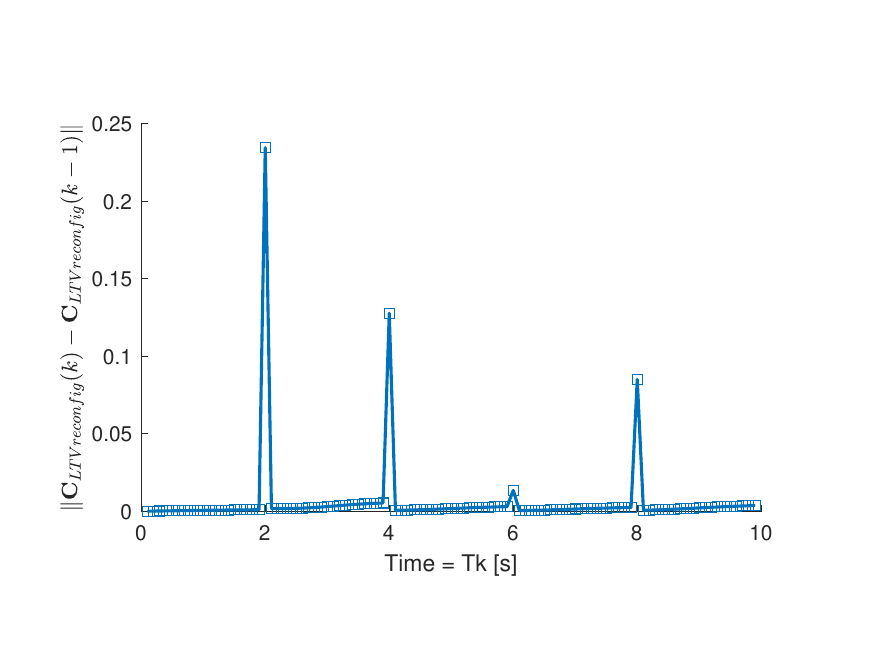}}
    \end{subfigmatrix}
    \caption{Reconfiguration scenarios of the spring-mass-damper system studied in this section.}
    \label{fig:reconfiguration}
\end{figure}

Developments in Robotics or Space Exploration create the need for safety-critical, accurate control systems. Controllers can only be as good as the system model allows. An optimal controller can have bad performance in the presence of strong epistemic uncertainty embodied on a too simplistic or ill-fit model. A few data-driven methods to fitting system models have recently come to light \cite{carvalho:2021:COSMICFastClosedformb,LTVModels,Pillonetto2023DeepNF}.
However, the impact of different modelling approaches on the end-objective of optimally controlling a highly perturbed and reconfiguring system has not been assessed, thus the trade-offs of modelling choice need to be determined to help engineering practice.
In this paper, we present the analysis of performance of the finite horizon Linear Quadratic Regulator (LQR) when applied to three discrete system models:
\begin{inparaenum}
    \item A reference linear time-invariant system (LTI),
    \item A linearized time-varying system (L-LTV), and
    \item A data-driven smoothed LTV model (DD-LTV) \cite{carvalho:2021:COSMICFastClosedformb}, \cite{LTVModels}.
\end{inparaenum}
% We tackle the problem of pointing a spacecraft to a comet during a fly-by, simulated according to the Comet Interceptor mission~\cite{esa:2019:AssessmentMissionIntercept}. The pointing requirement can be posed as an attitude guidance and control problem, aiming at keeping the comet within the field-of-view of the onboard scientific instruments as shown in Figure~\ref{fig:variation}(a). The fly-by is performed at high speed from a large distance to the comet, which makes meeting the requirement harder the closer the spacecraft gets to the target due to the need of significant torques and angular velocities (see Figure~\ref{fig:variation}(b)). Considering the linearized error dynamics w.r.t. the reference attitude, angular velocity, and torque, it becomes clear that the evolution is driven by the reference angular velocity, thus also having relevant variations throughout the scenario. However, it can be observed that this variation between samples is bounded, when considering a discrete-time setting as shown in Figure~\ref{fig:variation}(c). For optimal controller design purposes, nonlinear systems such as the error dynamics can benefit from being modelled as linear time-varying (LTV)~\cite{schoukens2019nonlinear,vanbeylen2013nonlinear}.
We tackle the problem of position tracking of a spring-mass-damper system. The tracking requirement can be posed as a position and velocity guidance and control problem, aiming at causing the mass to follow a desired trajectory. In different scenarios considered, the system gets perturbed or undergoes either an instantaneous, continuous, or mixed reconfiguration, where the parameters of the system change in a corresponding manner (see Figure~\ref{fig:reconfiguration}). It can be observed that the variation of the error dynamics of the systems between samples is bounded, when considering a discrete-time setting as shown for the mixed reconfiguration scenario in (see Figure~\ref{fig:variation}). For optimal controller design purposes, nonlinear systems such as the error dynamics can benefit from being modelled as linear time-varying (LTV)~\cite{schoukens2019nonlinear,vanbeylen2013nonlinear}.

% The identification of a system's behavior constitutes a problem in many areas of engineering~\cite{aastrom1971system}, from biological relations~\cite{crampin2006system} to physical systems' dynamics~\cite{agbabian1991system}, and it is the first step to any control design problem. It is imperative to have a system model that correctly represents its interaction with the environment and allows for more accurate predictions of its response to external stimuli. 

% Typically, this issue is addressed by considering previous knowledge about the system, applying first principles of physics and measuring whenever possible~\cite{Glad2013}. However, for more intricate systems, this approach fails because it may become impossible to measure all the parameters needed to characterize a behavior completely or the system is too complex to be modeled by simple equations~\cite{budiyono2011modeling}. In order to address these limitations, a data-driven approach to the system identification problem, where input-output data is used to find a model that describes a particular system, has become more common~\cite{brunton2019data}. Finding more efficient and comprehensive universal algorithms to deal with a large amount of data is essential to derive reasonable data-driven solutions and constitutes a pressing issue~\cite{cheng2015interplay}.

According to Lamnabhi-Lagarrigue et al.,~\cite{lamnabhi2017systems}, system modelling represents a significant cost in complex engineering projects, sometimes up to 50\% of the total cost, mainly due to the man-hours of the expert engineers dedicated to this task. Thus, as it is also stated, it is essential to create practical system identification tools that adapt to a wide range of problems and achieve a solution in a time-constrained setting. Such tools can be especially useful in a Space mission project that represents huge cost efforts for entire countries and agencies. The integration of Machine Learning (ML), and Guidance, Navigation and Control (GNC) can be helpful in this set of problems, from  building robust control frameworks that address parameter varying systems to applying verification and validation techniques to a system in a more efficient way.

\subsection{Related work}

 \subsubsection{Data-driven System Identification for LTV Systems}
 A significant body of research has focused on incorporating diverse forms of data into system identification, 
 aiming to improve model accuracy and adaptability. Methods that leverage data-driven techniques have been proposed 
 to identify linear time-varying (LTV) systems, as seen in~\cite{huang2021system, yin2021maximum, Pillonetto2023DeepNF}. 
 These approaches address the complexities associated with capturing the evolving dynamics of such systems and provide 
 flexible frameworks that are less reliant on rigid assumptions.
 
 Majji et al.~\cite{TVERA} developed a time-varying generalization of the classical eigensystem realization algorithm 
 (ERA)~\cite{ERA}. Additionally, the authors of~\cite{TVOKID} introduced a time-varying extension of the observer/Kalman 
 filter identification (OKID)~\cite{OKID}, improving numerical stability and reducing computational challenges 
 in the identification of lightly damped systems.
 
 \subsubsection{Control Design under System Uncertainty}
 Designing robust controllers for systems with uncertainty remains a critical area of study. 
 Efforts to integrate system identification with control synthesis can be found in~\cite{mohammadi2021convergence, de2019formulas, huang2022convex, rueda2021data}. 
 A notable trend is the development of data-driven model predictive control (MPC) 
 frameworks~\cite{berberich2020data, geromel2021sampled, rosolia2021robust} that provide guarantees of 
 robustness against model inaccuracies. Moreover, researchers have explored the feasibility of applying 
 these techniques to nonlinear systems~\cite{possieri2021iterative, guo2021data}, which are traditionally 
 challenging to model and control due to their complex dynamics.
 
 \subsubsection{Classical vs. Data-driven System Identification}
 A variety of LTV system identification methods have been proposed, ranging from classical approaches such as 
 TVERA~\cite{TVERA} and TVOKID~\cite{TVOKID} to more recent data-driven techniques like 
 COSMIC~\cite{carvalho:2021:COSMICFastClosedformb} and LTVModels~\cite{LTVModels}. Data-driven approaches, 
 particularly those employing regularization, have shown potential advantages in modeling smooth transitions in time-varying dynamics. 
 However, the impact of these methods on control performance, particularly in perturbed or reconfiguring systems, 
 has not been empirically evaluated to the same extent. 
 This paper aims to bridge this gap by comparing both classical and data-driven system identification 
 approaches and assessing their influence on control design performance in spring-mass-damper systems 
 undergoing various reconfiguration scenarios.

\subsection{Contributions}
This paper provides a practical analysis of machine learning techniques for system identification, 
focusing on their implications for optimal control design in perturbed and reconfiguring systems. 
Our key contributions are as follows:
\begin{itemize}
    \item We introduce a benchmark of LTV system identification methods, including LTVModels, TVERA, TVOKID, and COSMIC.
    \item We apply these methods to several perturbed and reconfiguring spring-mass-damper systems, 
    assessing the prediction accuracy and controller performance.
    \item We demonstrate that COSMIC-based system identification significantly improves performance in trajectory tracking 
    tasks compared to other classical LTV methods.
\end{itemize}

\section{Data-driven system identification}
\label{sec:DD-LTV}

We denote a discrete linear time-varying (LTV) system transition equation as
\begin{equation} \label{eq:sys_def}
    \vect{x}(k+1) = \vect{A}(k)\vect{x}(k) + \vect{B}(k) \vect{u}(k),
\end{equation}
where $k \in \{0, \ldots, N-1\}$, and $N+1$ is the total number of samples considered in a single trajectory. 
At a given time step $k$, the state vector is $\vect{x}(k) \in \Rbb{p}$, and the control input is $\vect{u}(k) \in \Rbb{q}$. 
The system parameters $\vect{A}(k) \in \Rbb{p \times p}$ and $\vect{B}(k) \in \Rbb{p \times q}$ are, respectively, the dynamics 
and control matrices defining the system's response.

\subsection{Fast Closed-form Identification from Large-scale Data for LTV Systems}
\label{sec:COSMIC}

Fast closed-form identification from large-scale data for LTV systems (COSMIC) is a data-driven LTV (DD-LTV) 
algorithm developed by Carvalho et al.~\cite{carvalho:2021:COSMICFastClosedformb}.
The COSMIC paper formulates an optimization problem where a data fidelity term is balanced with a regularization term 
that penalizes large variations in system dynamics between consecutive samples.
A closed-form solution is provided for this problem, which scales linearly with the number of time steps considered. 
Conditions necessary for obtaining a valid solution are also defined, and a preconditioning step is introduced to handle 
ill-conditioned data.

Assuming a dataset is collected with $L$ trajectories over $N+1$ time steps per trajectory, the state measurements 
are represented by the matrix $\vect{X}(k) \in \Rbb{p \times L}$, defined as 
\[
\vect{X}(k) =  \begin{bmatrix}
        \vect{x}_1(k) & \vect{x}_2(k) & \cdots & \vect{x}_L(k)
    \end{bmatrix},
\]
and the control inputs by $\vect{U}(k) \in \Rbb{q \times L}$ as 
\[
\vect{U}(k) =  \begin{bmatrix}
        \vect{u}_1(k) & \vect{u}_2(k) & \cdots & \vect{u}_L(k)
    \end{bmatrix}.
\]
The matrix $\vect{X'}(k)$ is defined as the state measurements at the next time step, i.e., $\vect{X'}(k) = \vect{X}(k+1)$.

\iffalse
\subsection{Fast closed-form identification from large-scale data for {LTV} systems}
Fast closed-form identification from large-scale data for {LTV} systems (COSMIC) is a DD-LTV algorithm developed by Carvalho et al.~\cite{carvalho:2021:COSMICFastClosedformb}.
It poses an optimiziation problem in which a data fidelity term is balanced with regularization penalizing for large system variability between two consecutive samples.
A closed form solution to this problem, that scales linearly with the number of time steps considered, is provided. Moreover, conditions that need to be satisfied in 
order to obtain a valid solution are defined. To solve the issue of ill-conditioned data, a preconditioning step is proposed.

Assume we collect a dataset with $L$ trajectories for the $N+1$ time steps per trajectory. Taking into account all the data available, we need to additionally define the matrices that contain the state measurements, $\vect{X}(k) \in \Rbb{p \times L}$ as $\vect{X}(k) =  \begin{bmatrix}
\vect{x}_1(k) & \vect{x}_2(k) & ... & \vect{x}_L(k)
\end{bmatrix}$ and the control data, $\vect{U}(k) \in \Rbb{q \times L}$ as $\vect{U}(k) =  \begin{bmatrix}
\vect{u}_1(k) & \vect{u}_2(k) & ... & \vect{u}_L(k)
\end{bmatrix},$ of all the $L$ different trajectories for the $k$-th instant of the trajectory. Moreover, we define $\vect{X'}(k)$ as $\vect{X'}(k) = \vect{X}(k+1)$.

\fi
To find the proper solution for system~\eqref{eq:sys_def}, we start by defining the optimization variable $\vect{C}(k) \in \Rbb{(p+q) \times p}$ as $\vect{C} (k) =  \begin{bmatrix}
    \vect{A}^T (k) \\ \vect{B}^T (k) \end{bmatrix}$, with $\vect{C} = \left[\vect{C}(k)\right]_{k=0}^{k=N-1}$, such that $\vect{C} \in  \Rbb{N(p+q) \times p}$.

    As such, the system identification problem is formulated
    \begin{equation}
        \begin{aligned}
            \underset{\vect{C}}{\text{minimize}}
            \hspace{0.2cm} f(\vect{C}) & :=  \frac{1}{2N} \| \vect{V} \vect{C} - \vect{X'}\|_F ^2 \\ & +   \frac{1}{2} \sum_{k=1}^{N-1} \| \lambda_k^{\onehalf} (\vect{C}(k) - \vect{C}(k-1))\|_F ^2,
        \end{aligned}
    \end{equation}
    where $\lambda_k > 0$ is a regularization constant for time step $k$, which serves the purpose of limiting system variability between two consecutive time instants.
    
According to \emph{Theorems 1 and 2} presented in the COSMIC paper, a valid solution to this problem is obtained under the condition that
the complete set of $\begin{bmatrix} \vect{x}^T_l(k) &  \vect{u}^T_l(k)\end{bmatrix}$ vectors within the dataset spans $\mathbb{R}^{p+q}$.

\subsection{Identification of LTV dynamical models with smooth or discontinuous time evolution by means of convex optimization}
\label{sec:LTVModels}
A DD-LTV algorithm similar to COSMIC was independently developed in \cite{LTVModels}. In section \emph{C} of the discussed paper, the authors define an optimization
problem minimizing data fidelity and non-squared L2 norm penalty for system variability:
\begin{equation}
    \mathop{minimize}_\vect{k} \|\vect{y} - \vect{\hat{y}}\|_{2}^{2} + \lambda \sum_t \| \vect{k}_{t+1} - \vect{k}_{t} \|_2 , 
\end{equation}
where $\vect{y}$ is a collection of true states, $\vect{\hat{y}}$ is a collection of predicted states, $\vect{k}_t = vect \left( \vect{C}(t) \right) =  vect \left( \begin{bmatrix}
\vect{A}^T (t) \\ \vect{B}^T (t) \end{bmatrix} \right)$, and $\lambda > 0$ is a constant adjusting the regularization strength.
This method identifies systems based on single trajectory data.

\subsection{Time-varying eigensystem realization algorithm}
\label{sec:TVERA}
In classical system identification we can find a time varying eigensystem realization algorithm (TVERA) \cite{TVERA}. It is based on
the utilization of data related to the free response of the system from non-zero initial conditions, and the response of a system to random inputs from zero
initial conditions, both for multiple experiments. Using the collected data, Hankel matrices are built. Singular value decomposition (SVD) is then applied
to obtain system matrices. Finally they are transformed to a common reference frame. This method does not involve smoothing but will be used for a comparison of
different LTV system identification methods as it represents a time-varying extension of a classical and well-studied system identification method ERA.

\subsection{Time-varying observer/Kalman fiter identification}
\label{sec:TVOKID}
Time-varying observer/Kalman filter identification (TVOKID) \cite{TVOKID} is another algorithm originating from classical system identification literature.
It eases the computations and numerical errors of applying the time-varying eigensystem realization algorithm to lightly damped systems by introducing an asymptotically stable observer.
TVOKID estimates Markov parameters that are later used by TVERA to identify system matrices, thus we refer to this joint method as TVOKID/TVERA. Just as TVERA, this
method does not involve smoothing but will be used for comparison of different LTV system identification algorithms as it is 
a time-varying extension of a classical and well-studied system identification method OKID.

\subsection{Modelled system}
\begin{figure}[htb]
    \begin{subfigmatrix}{2}
        \subfigure[Schematic of a typical spring-mass-damper system. During instantaneous reconfiguration,
        coefficients $m$, $C_s$, and $C_d$ undergo discontinuous variations in time.\label{fig:smd}]
        {\includegraphics[width=0.49\linewidth]{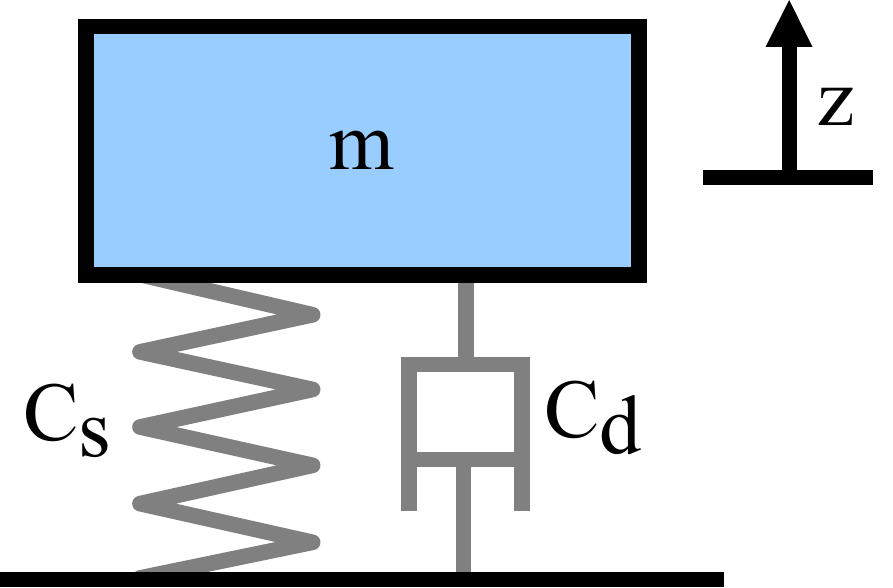}}
        \subfigure[Variation of the dynamics of the (linearized) LTV system \eqref{eq:LTV-system}.\label{fig:LTV-C}]
        {\includegraphics[width=0.49\linewidth, trim=25 40 45 40,clip]{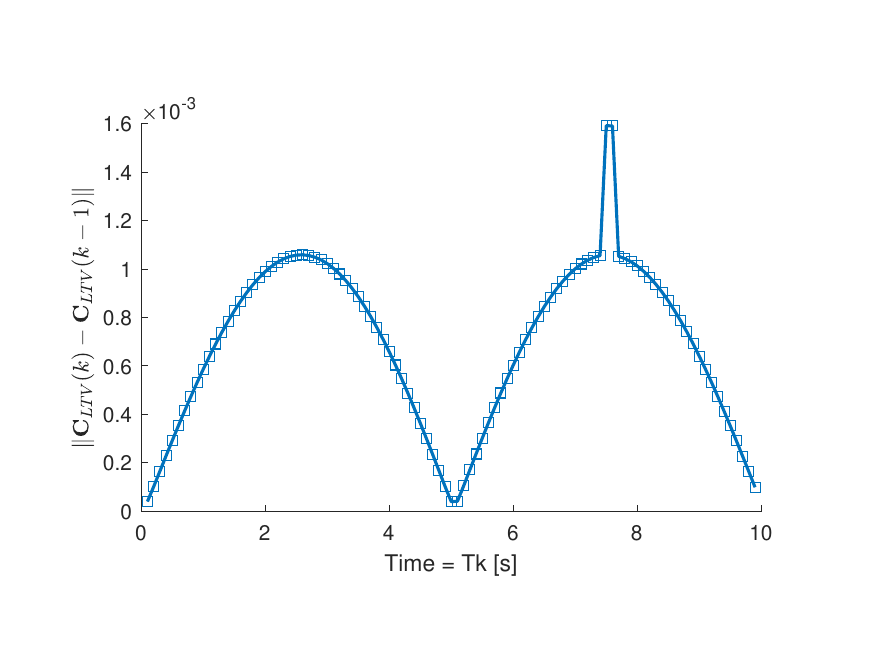}}
    \end{subfigmatrix}
    \caption{The spring-mass-damper system studied in this section.}
\end{figure}

Let us consider a classical spring-mass-damper system (see figure \ref{fig:smd} for a graphical representation and \cite[Section 1.2.3]{khalil:2002:NonlinearSystems} for more details) as
\begin{equation}\label{eq:smd}
    m(t)\Ddot{z}(t) = -c_d(t,\Dot{z}(t)) - C_s(t) z(t) + ext(u(t))
\end{equation}
with $z(t)$ as the mass position in relation to the resting point at $z = 0$, $C_s(t)$ is the spring constant,
$c_d(t,\Dot{z})$ represents the damping effect, and $ext(u(t))$ represents the effects of external forces $u(t)$ on the system. For a linear system, the damping effect can be written as $c_d(t,\Dot{z})=C_d(t)\Dot{z}(t)$, where $C_d(t)$ is the damping coefficient, and typically $ext(u(t))=u(t)$.

If we denote $\vect{x} = \begin{bmatrix}
        z &
        \Dot{z}
    \end{bmatrix}^T=\begin{bmatrix}
        x_1 &
        x_2
    \end{bmatrix}^T$ as the state, the continuous time state space system can be addressed as

\begin{equation}\label{eq:LTV:cont}
    \Dot{\vect{x}}(t) =  \underbrace{\begin{bmatrix}
            0                    & 1                    \\
            -\frac{C_s(t)}{m(t)} & -\frac{C_d(t)}{m(t)} \\
        \end{bmatrix}}_\text{$\vect{A}_c$} \vect{x}(t) +  \underbrace{\begin{bmatrix}
            0              \\
            \frac{1}{m(t)} \\
        \end{bmatrix}}_\text{$\vect{B}_c$} u(t).
\end{equation}
where we can approach this system as a LTV (continuous reconfiguration), by varying parameters $C_s(t)$ and $C_d(t)$, and keeping $m(t)=m$ constant. For this particular case, we defined
\begin{equation} \label{eq:LTV-parameters}
    \begin{cases}
        C_{s}(t) = \cos \left( 1.5 \omega_0 t + \frac{\pi}{4} \right) ^2 C_s \\
        C_{d}(t)= \left[ 1.5 +  \cos \left(\omega_0 t \right) \right ] C_d
    \end{cases}.
\end{equation}

To represent this scenario as a discrete-time linear system, a discretization of the problem was performed, assuming piecewise-constant inputs $u(k)$ and parameters $C_{s}(k),C_{d}(k)$, with $m$ and $\Delta t$ constants and $t=k\Delta t$, resulting in a linearized discrete version
\begin{equation} \label{eq:LTV-system}
    \vect{x}(k+1) =  \underbrace{e^{\vect{A}_c(k) \Delta t}}_\text{$\vect{A} (k)$}  \vect{x}(k) +  \underbrace{\vect{A}_c^{-1}(k) (e^{\vect{A}_c(k) \Delta t} - \vect{I}) \vect{B}_c}_\text{$\vect{B}(k)$} u(k).
\end{equation}
For this system, the evolution of $\vect{C}(k)-\vect{C}(k-1)$, i.e., the variation of the dynamics, can be found in figure
\ref{fig:LTV-C}.

To further demonstrate the performance of COSMIC and its impact on control design, recall \eqref{eq:smd}. A new, nonlinear (NL) system is defined considering a nonlinear damping $c_d(t,\Dot{z}(t))=C_d(t)\Dot{z}(t)+C_{d_3}\Dot{z}^3(t)$ \cite{elliott:2015:NonlinearDampingQuasilinear} and a saturation of the input $ext(u(t))=\mathrm{sat}_U\!(u(t))$, resulting in
\begin{equation}\label{eq:NL:cont}
    \Dot{\vect{x}}(t) =  \begin{bmatrix}
        0                 & 1                                 \\
        -\frac{C_s(t)}{m} & -\frac{C_d(t)+C_{d_3}x_2^2(t)}{m} \\
    \end{bmatrix} \vect{x}(t) +  \begin{bmatrix}
        0           \\
        \frac{1}{m} \\
    \end{bmatrix} \mathrm{sat}_U\!(u(t)).
\end{equation}
which linearization and discretization is also \eqref{eq:LTV-system}.

Additionally, an impulsive stochastic disturbance $d(x_1(t))$ dependent on the distance of the mass to a particular point ($x_1=2$) is added to $x_2(t)$ for the nonlinear perturbed system (NLD)
\begin{equation}\label{eq:NLD:cont}
    \begin{aligned}
        \Dot{\vect{x}}(t) & =  \begin{bmatrix}
                                   0                 & 1                                 \\
                                   -\frac{C_s(t)}{m} & -\frac{C_d(t)+C_{d_3}x_2^2(t)}{m} \\
                               \end{bmatrix} \vect{x}(t) \\
                          & +  \begin{bmatrix}
                                   0           \\
                                   \frac{1}{m} \\
                               \end{bmatrix} \mathrm{sat}_U\!(u(t))
        +\begin{bmatrix}
             0         \\
             d(x_1(t)) \\
         \end{bmatrix}
        .
    \end{aligned}
\end{equation}
Again, the linear discrete version is given by \eqref{eq:LTV-system}.

To simulate the scenario of instantaneous reconfiguration (LTV instantaneous reconfiguration), the parameters $C_s(t)$, $C_d(t)$, and $m(t)$ are constant within predefined two-second-long time frames and change instantaneously between them. Moreover, each parameter
change is accompanied by a stochastic disturbance of the system's velocity $d_v(t)$, which attains non-zero values only when transitioning to a new two-second-long time frame.
\begin{equation}\label{eq:inst_reconfig:cont}
    \begin{aligned}
        \Dot{\vect{x}}(t) & =  \begin{bmatrix}
                                   0                    & 1                    \\
                                   -\frac{C_s(i)}{m(i)} & -\frac{C_d(i)}{m(i)} \\
                               \end{bmatrix} \vect{x}(t) \\
                          & +  \begin{bmatrix}
                                   0              \\
                                   \frac{1}{m(i)} \\
                               \end{bmatrix} u(t)
        +\begin{bmatrix}
             0      \\
             d_v(t) \\
         \end{bmatrix}
        ,
    \end{aligned}
\end{equation}
where $i=\left \lfloor{\frac{t}{2}}\right \rfloor$ is the index of a current two-second-long time frame. As in the case of previous systems, the linear discrete version is given by \eqref{eq:LTV-system}.

Finally, to consider the scenario of a mixed reconfiguration (LTV reconfiguration), the parameters $C_s(t)$, and $C_d(t)$ follow the time-dependent variations as defined in \eqref{eq:LTV-parameters} inside every two-second-long time frame. Between these two-second-long intervals, $m$, $C_s$, and $C_d$ change instantaneously. Once again each instantaneous parameter
change is accompanied by a stochastic disturbance of the system's velocity $d_v(t)$, which attains non-zero values only when transitioning to a new two-second-long time frame.
\begin{equation}\label{eq:reconfig:cont}
    \begin{aligned}
        \Dot{\vect{x}}(t) & =  \begin{bmatrix}
                                   0                    & 1                    \\
                                   -\frac{C_s(t)}{m(i)} & -\frac{C_d(t)}{m(i)} \\
                               \end{bmatrix} \vect{x}(t) \\
                          & +  \begin{bmatrix}
                                   0              \\
                                   \frac{1}{m(i)} \\
                               \end{bmatrix} u(t)
        +\begin{bmatrix}
             0      \\
             d_v(t) \\
         \end{bmatrix}
        ,
    \end{aligned}
\end{equation}
where
\begin{equation} \label{eq:LTV-parameters-reconfig}
    \begin{cases}
        C_{s}(t) = \cos \left( 1.5 \omega_0 t + \frac{\pi}{4} \right) ^2 C_s(i) \\
        C_{d}(t)= \left[ 1.5 +  \cos \left(\omega_0 t \right) \right ] C_d(i)
    \end{cases},
\end{equation}
and $i=\left \lfloor{\frac{t}{2}}\right \rfloor$ is the index of a current two-second-long time frame. As previously, the linear discrete version is given by \eqref{eq:LTV-system}.

\subsection{Experimental setup}
In order to correctly identify the considered systems, it is important to properly excite them during data collection. For this purpose, the system of interest is setup in an open loop
with a disturbed chirp input signal:
\begin{equation} \label{eq:chirp}
    u(t) = A \sin \left( \frac{\omega_1 - \omega_0}{T}t^2 + \omega_0 t + \phi \right) + d_c(t),
\end{equation}
where $\omega_1$, $\omega_0$ are the maximal and minimal frequencies of the input signal, $T$ is the duration of the whole simulation, $\phi$ is phase, $d_c(t)$ is a stochastic signal with mean 0 and variance $\sigma^2$, and $t$ is time.

\begin{figure}[htb]
    \centering
    \includegraphics[width=0.5\textwidth]{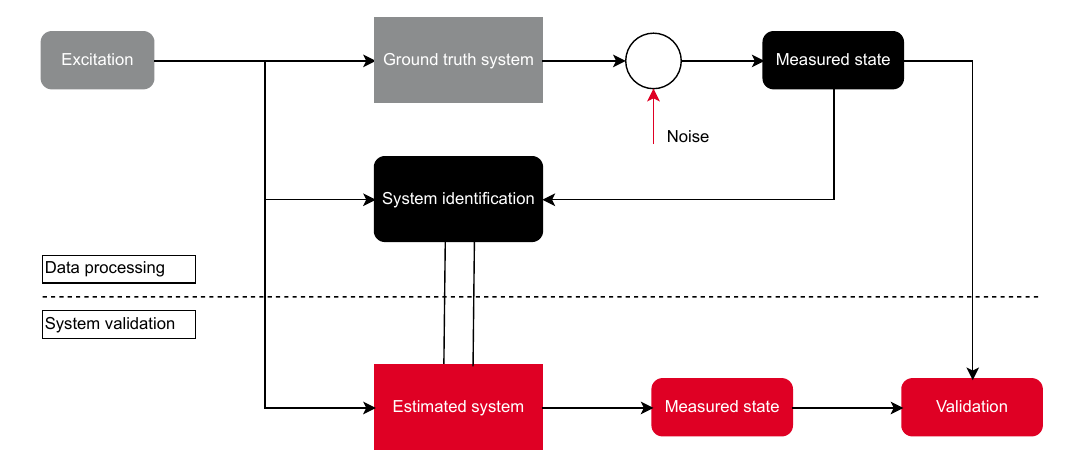}
    \caption{Schematic representing the validation procedure for hyperparameter tuning and performance evaluation. Noise is injected to the state measurements
    only when collecting the training data for system identification. The validation and testing datasets contain true state measurements to ensure proper validation.}
    \label{fig:validation}
\end{figure}

Given that the considered system identification methods require hyperparameter tuning, the data collection process is split into three parts so that training, validation, and testing datasets are created. Between these datasets, the frequencies $\omega_1$ and $\omega_0$ are varied.
As opposed to the training dataset, the validation and testing datasets contain true state measurement, without measurement noise. Moreover, they are supplied with data collected with $u(t)=0$ and non-zero initial conditions. The validation dataset is used to tune hyperparameters and the testing dataset to obtain an ubiased estimate of the resulting model's performance.
A flowchart representing this process is presented in Figure~\ref{fig:validation}.

The process of hyperparameter tuning is based on evaluating how well a given model performs in the task of predicting future states of the system over a certain time horizon, given initial conditions and control inputs over that horizon.
For this purpose a RMSE averaged over the validation trajectories is defined, further referred to as the trajectory prediction loss:
\begin{equation} \label{eq:trajectory-loss}
    L(\vect{\beta}) = \frac{1}{\vect{|S|}} \sum_{l \in \vect{S}} \sqrt{\frac{1}{N} \sum_{k=1}^{N} \sum_{m=1}^{p} \left( \hat{x}_{k, m}^{(l)}(\vect{\beta}) - x_{k, m}^{(l)} \right)^2},
\end{equation}
where $\vect{\beta}$ corresponds to hyperparameters, $\hat{x}_{k, m}^{(l)}(\vect{\beta})$ is the $m$-th element of the $p$-dimensional state vector predicted from initial conditions by the identified system at time step $k$ for trajectory $l$, $x_{k, m}^{(l)}$ is a corresponding true state of the system, and $\vect{S}$ represents a set of trajectories.
To find the best hyperparameters, one selects a model that results in the smallest trajectory prediction loss for the set of validation trajectories.

\subsection{Comparison of LTV system identification methods}
Using the presented experimental setup, we first collect the datasets, imposing the requirements of the methods on the properties
of the corresponding training datasets. Then we find the best hyperparameters of the described LTV system identification methods for all considered scenarios and estimate system models.
The final performance of these algorithm is assessed by computing the trajectory prediction loss \eqref{eq:trajectory-loss} on the testing dataset.

Given that the LTVModels method identifies systems using a single trajectory, we conducted the process of system identification with COSMIC twice,
using a single and multiple trajectories, so that the resulting benchmark of methods is more fair.

The analysis of the trajectory prediction losses obtained by identifying the systems with different methods presented in Table~\ref{table:ltv-comparison} clearly shows the advantage of the DD-LTV model. The multi-trajectory version of COSMIC method, for all the systems except for NL \eqref{eq:NL:cont}, resulted in the most accurate predictions of trajectories from initial conditions.
In the case of the NL system, the best performance was observed for a single trajectory version of COSMIC and LTVModels simultaneously. The non-smoothed methods, i.e., TVERA and TVERA/TVOKID
did not beat the smoothed models in any of the considered scenarios.

\begin{table}[!ht]
    \centering
    \tiny
    \caption{Comparison of trajectory prediction loss on the testing dataset for different LTV system identification methods.}
    \label{table:ltv-comparison}
        \begin{tabular}{>{\hspace{0pt}}m{0.11\linewidth}>{\hspace{0pt}}m{0.067\linewidth}>{\centering\hspace{0pt}}m{0.09\linewidth}>{\centering\hspace{0pt}}m{0.09\linewidth}>{\centering\hspace{0pt}}m{0.09\linewidth}>{\centering\hspace{0pt}}m{0.09\linewidth}>{\centering\arraybackslash\hspace{0pt}}m{0.09\linewidth}} 
        \hline
        \textbf{System} & \textbf{RMSE} & \textbf{TVERA} & \textbf{TVERA/ TVOKID} & \textbf{LTVModels} & \textbf{COSMIC single trajectory} & \textbf{COSMIC}                                       \\ 
        \hline
        LTV \eqref{eq:LTV:cont}            & mean          & 0.71           & 0.52                  & 0.37               & 0.36                              & \textbf{0.01}                                         \\
                                            & std           & 0.52           & 0.37                  & 0.21               & 0.18                              & \textbf{0.01}                                         \\ 
        \hline
        NL \eqref{eq:NL:cont}              & mean          & 1.15           & 1.07                  & \textbf{0.32}      & \textbf{0.32}                     & 0.35                                                  \\
                                        & std           & 0.91           & 0.55                  & \textbf{0.21}      & \textbf{0.21}                     & 0.23                                                  \\ 
        \hline
        NLD \eqref{eq:NLD:cont}             & mean          & 5.95           & 2.61                  & 2.41               & 2.41                              & \textbf{2.20}                                         \\
                                            & std           & 2.21           & \textbf{0.90}         & 1.04               & 1.16                              & 0.94                                                  \\ 
        \hline
        LTV                                & mean          & 0.61           & 0.47                  & 2.14               & 2.50                              & \textbf{0.02}                                         \\
        inst                               & std           & 0.41           & 0.33                  & 1.13               & 1.51                              & \textbf{0.01}                                         \\
        reconfig \eqref{eq:inst_reconfig:cont}        &               &                &                       &                    &                                   &                                                       \\ 
        \hline
        LTV                                      & mean          & 0.65           & 0.31                  & 1.41               & 1.36                              & \textbf{0.01}                                         \\
        reconfig \eqref{eq:reconfig:cont}        & std           & 0.52           & 0.23                  & 0.71               & 0.76                              & \textbf{0.00}                                         \\
        \hline
    \end{tabular}
\end{table}

\begin{figure}[!htb]
    \begin{subfigmatrix}{2}
        \subfigure[LTV scenario.\label{fig:ecdf-ltv}]
        {\includegraphics[width=\linewidth,trim=0 0 25 20,clip]{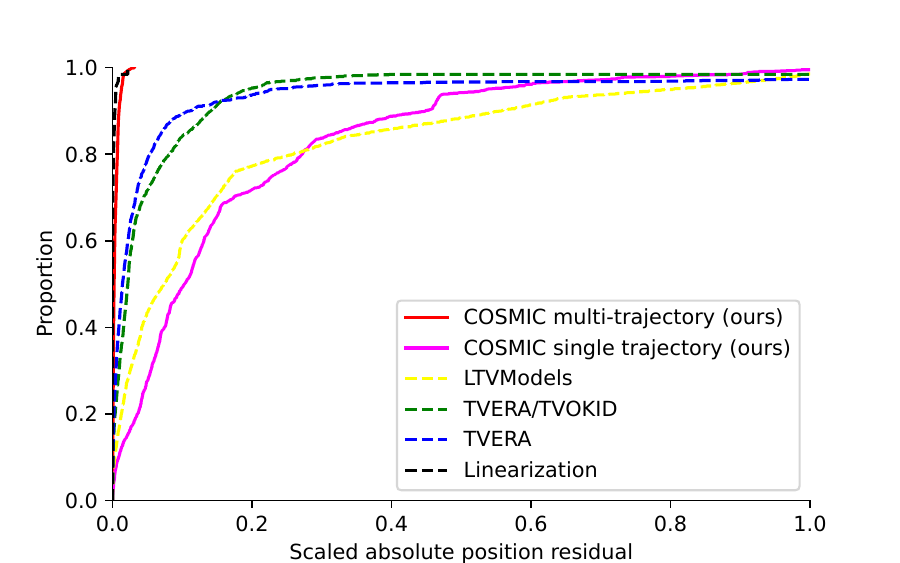}}
        \subfigure[LTV reconfiguration scenario.\label{fig:ecdf-reconfig}]
        {\includegraphics[width=\linewidth,trim=0 0 25 20,clip]{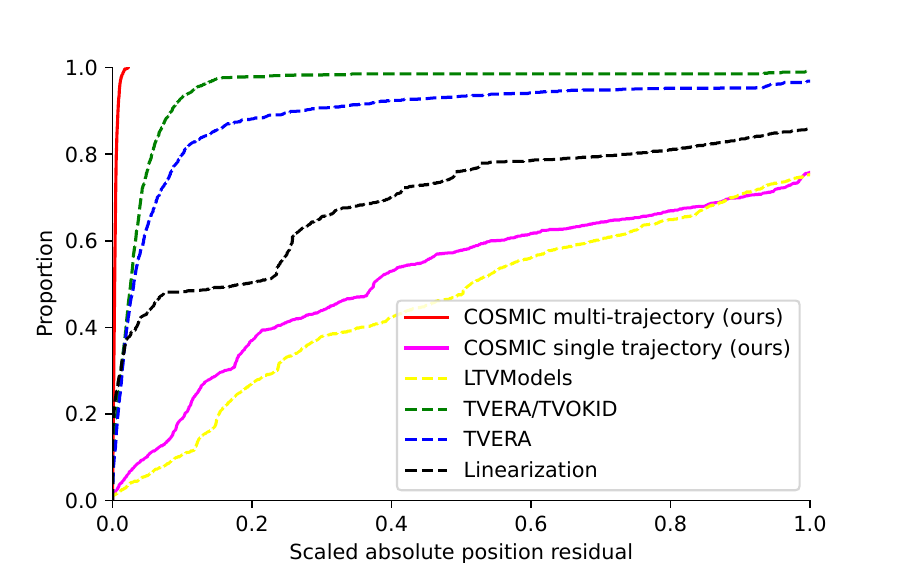}}
    \end{subfigmatrix}
    \caption{ECDFs of relative absolute position residuals in the task of testing trajectory prediction from initial conditions. Yellow and magenta lines represent a similar DD-LTV optimization problem thus they follow the same trend.}
    \label{fig:ecdf}
\end{figure}

To facilitate the understanding of how these system identification methods perform, for the LTV and LTV mixed reconfiguration systems, the absolute values of position residuals obtained for a
given trajectory were scaled by the mean absolute position of this trajectory. Then their empirical cumulative distribution functions (ECDFs) were plotted (Figure~\ref{fig:ecdf}).
This visualization clearly shows the advantages of the DD-LTV approach to system identification over the classical methods. The resulting model of the system enables state propagation
with linearization-like accuracy for the LTV system (Figure~\ref{fig:ecdf-ltv}) or significantly more accurate than the corresponding linearization (Figure~\ref{fig:ecdf-reconfig}) in the case of reconfiguration.
Given this good performance, we further investigate the problem of controller design based on the models estimated by COSMIC.

\section{Controller design}
\label{chapter:results}

\begin{figure}[htb]
    \centering
    \includegraphics[width=0.5\textwidth]{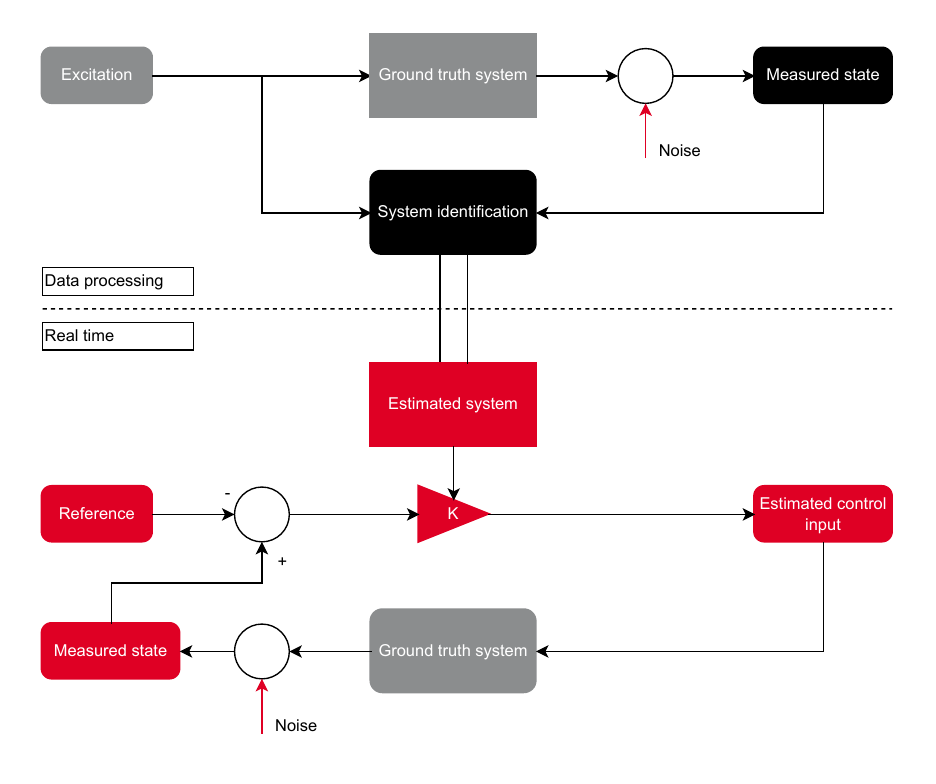}
    \caption{Schematic representing the procedure of testing the identified systems in the task of reference tracking.}
    \label{fig:real-time}
\end{figure}

The best COSMIC model obtained in the process of hyperparameter tuning is used for controller design. An optimal control law is obtained for the identified system and applied in a closed loop to the ground truth system in the task of reference trajectory tracking. This process is presented in Figure~\ref{fig:real-time}.

\begin{figure}[!htb]
    \centering
    \begin{subfigmatrix}{2}
        \subfigure[Hyperparameter $\lambda$ vs COSMIC cost.\label{fig:lambda-cost}]
        {\includegraphics[width=\linewidth,trim=25 40 45 40,clip]{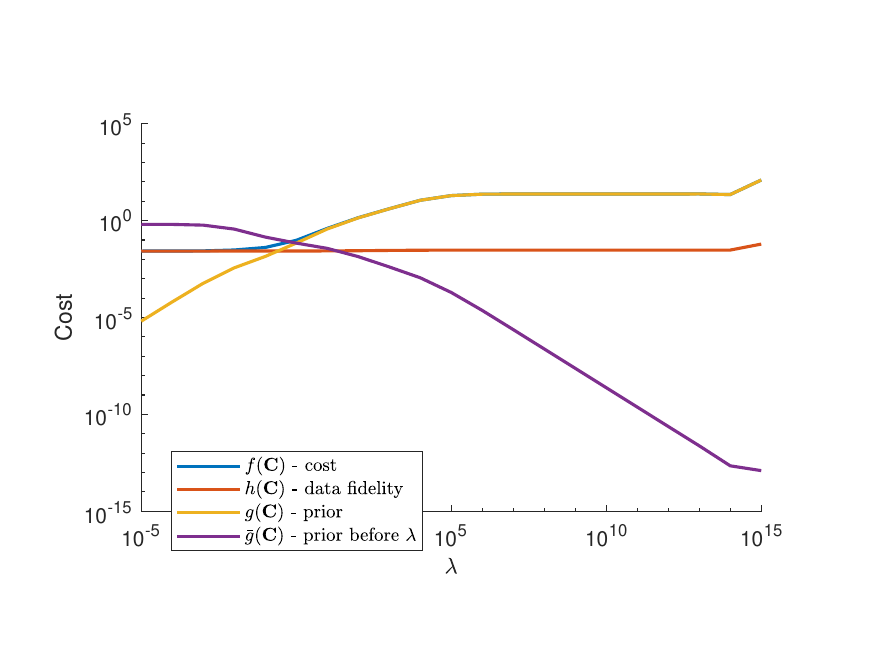}}
        \subfigure[Hyperparameter $\lambda$ vs reference tracking RMSE.\label{fig:lambda-control}]
        {\includegraphics[width=\linewidth,trim=25 40 45 40,clip]{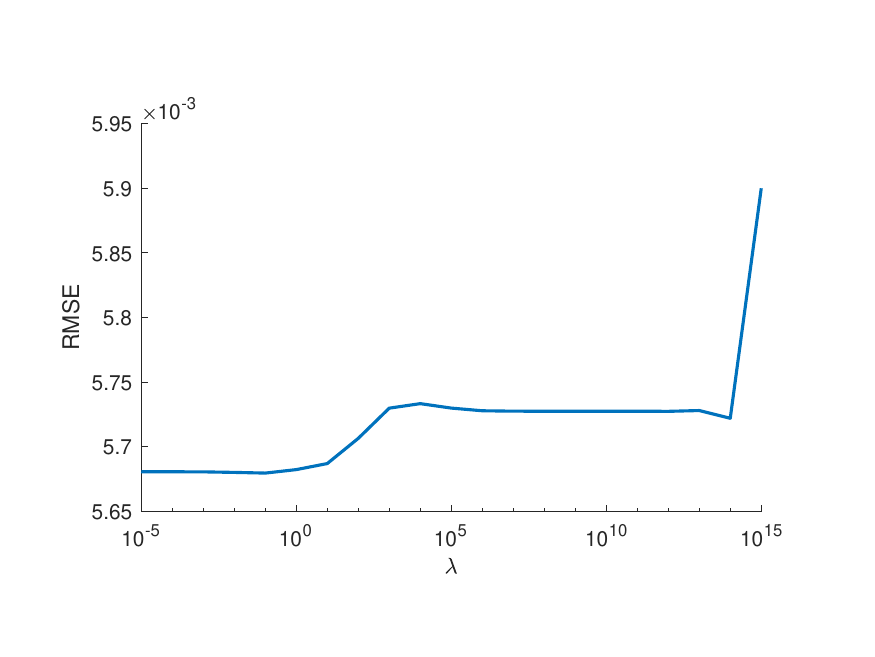}}
    \end{subfigmatrix}
    \caption{Effects of $\lambda$ on the COSMIC estimated model of the system and resulting controller performance for the LTV scenario. Increasing the value of this hyperparameter reduces the
    system variability between consecutive time samples, resulting in a decreased \emph{prior before $\lambda$} term. This change affects the reference tracking performance of
    the LQR controller designed based on the estimated model.}
    \label{fig:lambda-effect}
\end{figure}

To follow up on the system identification from data, we derive an optimal control law for the LTV estimated model based on a dynamic programming setup, as detailed in~\cite{zak2003systems} for discrete-time LTV LQR.
We address the issue from the last point of the optimal solution and state that the minimal cost for this time step
is $J_N = \frac{1}{2} \vect{x}^T_N \vect{H} \vect{x}_N$. Algorithm~\ref{alg:controller} summarizes this approach.
\begin{algorithm}[!htb]
    \caption{Dynamic Programming for LTV controller design}\label{alg:controller}
    \begin{algorithmic}[1]
        \Require{$\vect{H}$, $\vect{Q}$, $\vect{R}$, $\vect{\vect{A}}$, $\vect{\vect{B}}$}
        \Ensure{$\vect{K}$}

        \Statex

        \State $\vect{P}_{N} \gets \vect{H}$

        \
        \For {$k \in  [N-1,...,0]$}
        \State $\vect{K}_{k} \gets \left( \vect{R}_k + \vect{B}^T(k) \vect{P}_{k+1} \vect{B}(k) \right)^{-1} \vect{B}^T(k) \vect{P}_{k+1} \vect{A}(k)$
        \State $\vect{P}_k \gets \vect{Q}_k + \vect{K}_k^T \vect{R}_{k} \vect{K}_k + \left( \vect{A}(k) - \vect{B}(k) \vect{K}_k \right)^T \vect{P}_{k+1} \left( \vect{A}(k) - \vect{B}(k) \vect{K}_k \right)$

        \EndFor

        \
        \Return $\vect{K}$

    \end{algorithmic}
\end{algorithm}

\begin{figure*}[htb]
    \centering
    \includegraphics[width=0.3\textwidth,trim=25 15 45 25,clip]{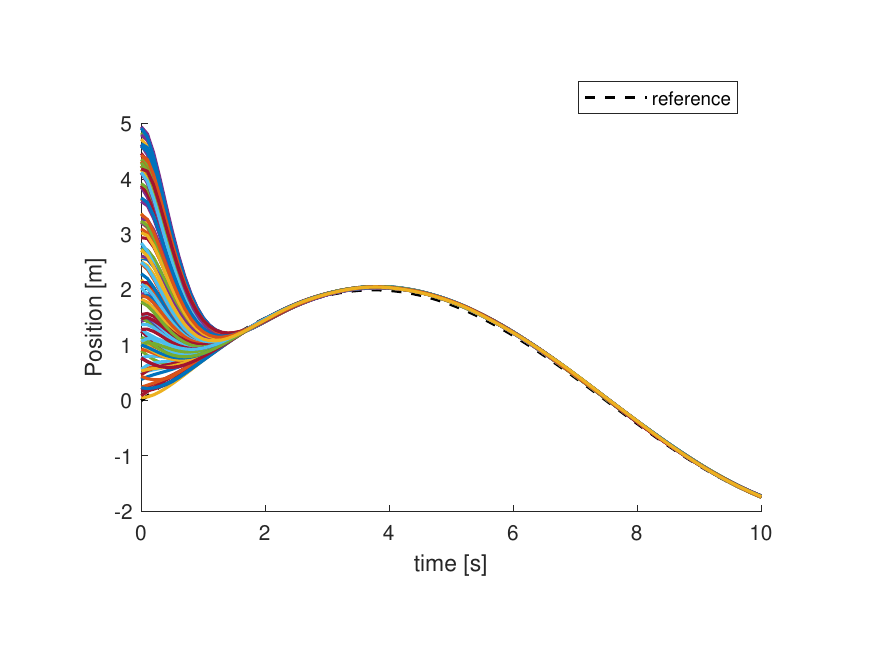}
    \includegraphics[width=0.3\textwidth,trim=25 15 45 25,clip]{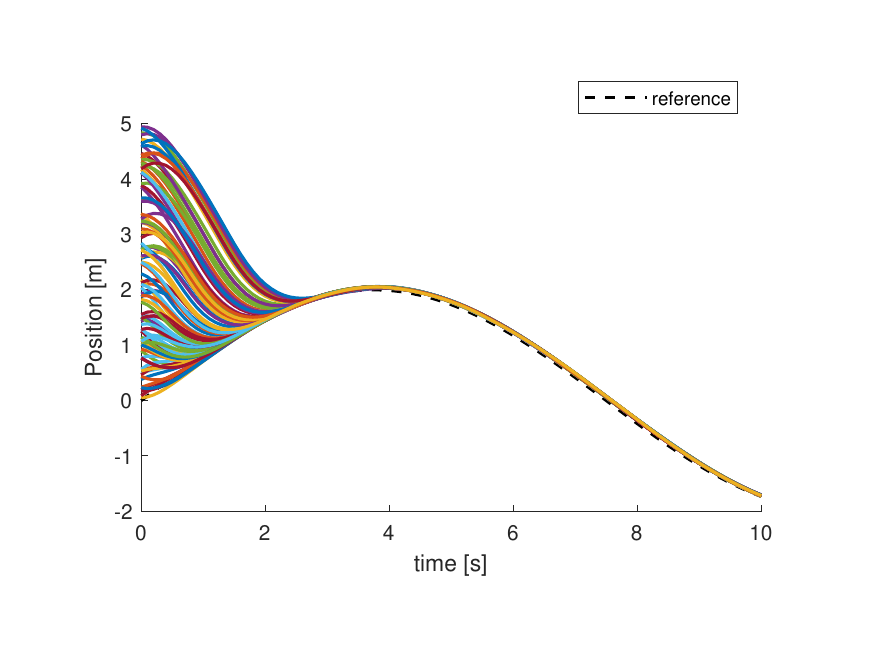}
    \includegraphics[width=0.3\textwidth,trim=25 15 45 25,clip]{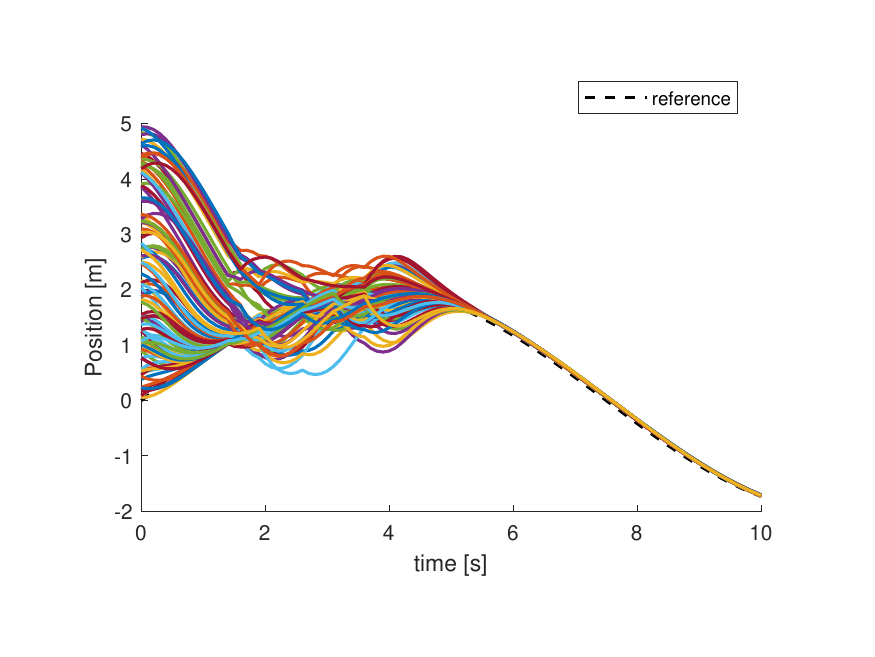}
    \includegraphics[width=0.3\textwidth,trim=25 15 45 25,clip]{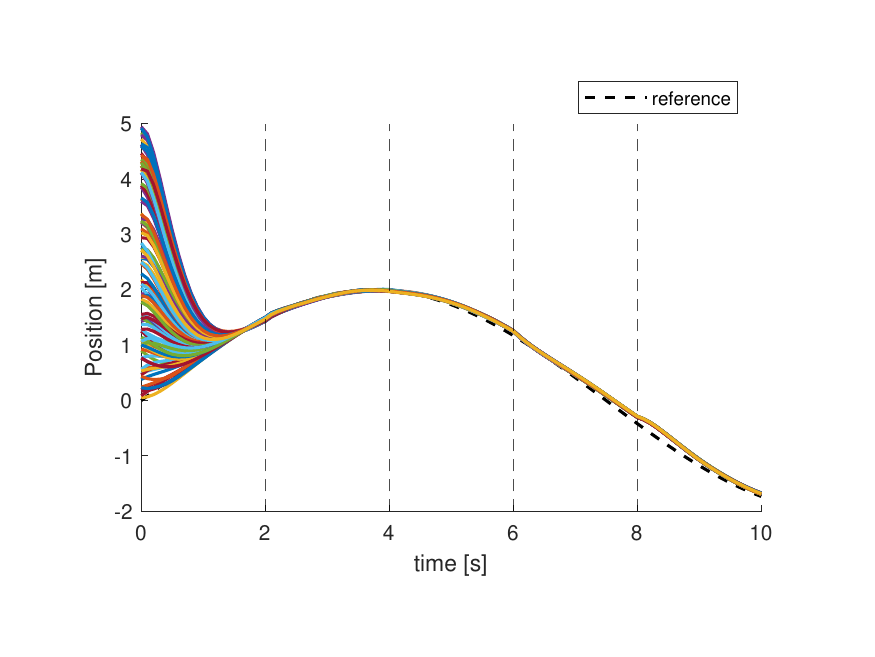}
    \includegraphics[width=0.3\textwidth,trim=25 15 45 25,clip]{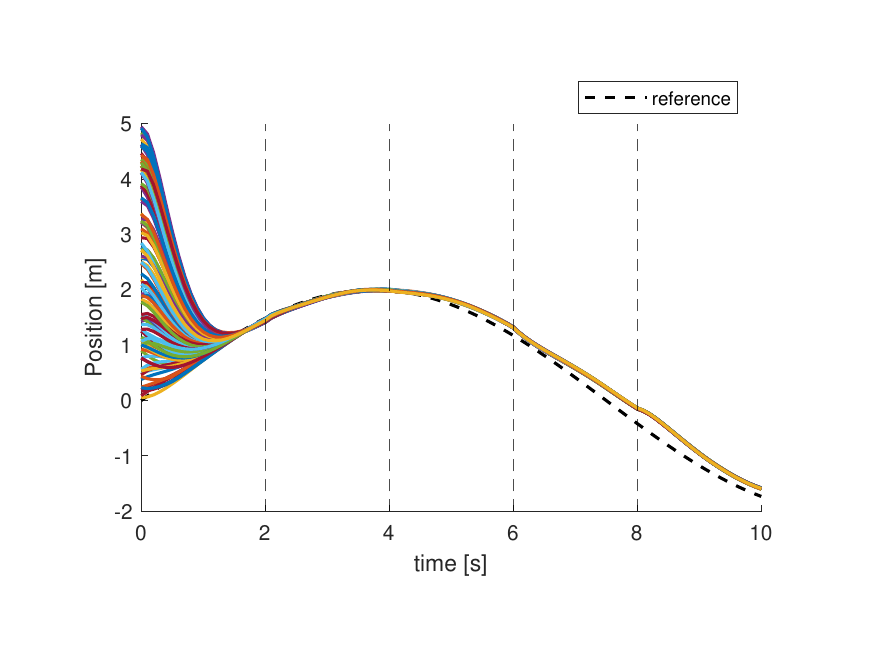}
    \caption{System position response from different initial conditions. The designed controller is able to follow the reference and respond quickly to different initial conditions. From left to right and top to bottom: LTV, NL, NLD, LTV instantaneous reconfiguration, LTV reconfiguration systems.
    The vertical dashed lines correspond to the times at which the systems underwent instantaneous reconfiguration.}
    \label{fig:init-cond}
\end{figure*}

For the design phase of the controller, we develop a tuning framework taking into account the LQR needs, keeping matrices $\vect{Q}_k$ and $\vect{R}_k$ constants throughout the trajectory and depending only on the elements from their diagonals, entailing that the most suiting design parameters are $\frac{q_v}{q_x} = 10^{-1}$, $\frac{r}{q_x} = 10^{-3}$ and $q_x = 1$. Figure~\ref{fig:init-cond} shows each system response to different inputs when controlled by the LTV LQR drawn from the estimated model. It is clear that the controller designed from the COSMIC estimated model can follow the reference from different initial conditions and it is a good solution to the control challenge.

After performing a statistical analysis of the tracking error, we can confidently say that the estimated model is a good approximation of the system, as shown by the close values of the statistical metrics in Table~\ref{table:stats-control-error}. Moreover, it allows for a better controller synthesis than previously used techniques, performing better than the LTI LQR, presenting a smaller mean and sum of squared error, once again from Table~\ref{table:stats-control-error}.

Thus, the model estimated by COSMIC is a good approximation of the spring-mass-damper system and the dynamic programming strategy worked well for the problem posed. Hence, this system identification and controller design framework is validated and we can further test it in more complex environments.

\begin{table}[ht!]
    \caption{Statistical comparison of the controller design methods. Mean, standard deviation, and RMSE relative to the absolute reference tracking errors for the optimal controllers considered.}
    \vspace{0.1cm}
    \begin{center}
        \begin{tabular}{llccc}
            \hline
            \textbf{System} & \textbf{Controller} & \textbf{Mean} & $\mathbf{\sigma}$ & \textbf{RMSE} \\
            \hline
            \input{errors.tex}
        \end{tabular}
        \label{table:stats-control-error}
    \end{center}
\end{table}

\section{CONCLUSIONS}

In this paper we evaluated different formulations of a smooth Linear Time Variant system learned from data (DD-LTV)
with state of the art Linear Time Variant (LTV) system identification methods in five spring mass damper scenarios, including
nonlinearly damped, highly perturbed, and reconfiguring systems whose parameters were undergoing continuous and/or discontinuous changes in time.
The analysis of residuals in the task of trajectory prediction from initial conditions showed the superiority of the DD-LTV approaches
over the classical LTV system identification methods in all the scenarios considered in the presented benchmark.

We further analyzed the best DD-LTV approach identified by this benchmark, i.e., \emph{fast closed-form identification from large-scale data for {LTV} systems (COSMIC)}
in the task of design of an optimal LTV LQR controller for reference trajectory tracking, based on the estimated models
of the considered system. We compared the performance of this controller, in terms of tracking errors,
against the ones obtained for Linear Time Invariant (LTI) and linearized Linear Time Variant (L-LTV) system models.
The DD-LTV approach resulted in better tracking performance than the LTI, and comparable performance to the L-LTV,
proving it to be a suitable system identification method for the purpose of optimal control.

Both the analysis of state propagation accuracy and trajectory tracking performance showed that the LTV state-space
representation is suitable for modelling nonlinear, highly perturbed, and reconfiguring systems. The optimal choice of system matrices enables reasonably
accurate state propagation as well as design of well-performing controllers for the task of trajectory tracking.

\bibliographystyle{IEEEtran}
\bibliography{autosam}
\end{document}

%% file: errors.tex
LTV \eqref{eq:LTV:cont}	&	 \textbf{COSMIC}	&	 \textbf{0.1733} 	&	 \textbf{0.0538} 	&	 \textbf{0.0057} \\
LTV	&	 Linearization	&	 0.1740 	&	 0.0540 	&	 \textbf{0.0057} \\
LTV	&	 Time-invariant	&	 0.1794 	&	 0.0546 	&	 0.0058 \\
\hline
NL \eqref{eq:NL:cont}	&	 \textbf{COSMIC}	&	 \textbf{0.2569} 	&	 0.0674 	&	 \textbf{0.0074} \\
NL	&	 Linearization	&	 0.2593 	&	 \textbf{0.0673} 	&	 \textbf{0.0074} \\
NL	&	 Time-invariant	&	 0.2625 	&	 0.0674 	&	 \textbf{0.0074} \\
\hline
NLD \eqref{eq:NLD:cont}	&	 \textbf{COSMIC}	&	 \textbf{0.3229} 	&	 \textbf{0.0668} 	&	 \textbf{0.0076} \\
NLD	&	 Linearization	&	 0.3248 	&	 0.0669 	&	 \textbf{0.0076} \\
NLD	&	 Time-invariant	&	 0.3267 	&	 0.0669 	&	 \textbf{0.0076} \\
\hline
LTV inst reconfig \eqref{eq:inst_reconfig:cont}	&	 \textbf{COSMIC}	&	 0.1905 	&	 \textbf{0.0538} 	&	 \textbf{0.0057} \\
LTV inst reconfig	&	 Linearization	&	 \textbf{0.1903} 	&	 0.0539 	&	 \textbf{0.0057} \\
LTV inst reconfig	&	 Time-invariant	&	 13.9992 	&	 3.4370 	&	 0.3683 \\
\hline
LTV reconfig \eqref{eq:reconfig:cont}	&	 \textbf{COSMIC}	&	  0.2442 	&	 \textbf{0.0528} 	&	 \textbf{0.0058} \\
LTV reconfig	&	 Linearization	&	 \textbf{0.2423} 	&	 0.0529 	&	 \textbf{0.0058} \\
LTV reconfig	&	 Time-invariant	&	 7.7488 	&	 1.8094 	&	 0.1957 \\
\hline